\documentclass[11pt,preprint]{aastex}

\begin{document}

\title{Limits on Collisional Dark Matter from Elliptical Galaxies in Clusters}

\author{Oleg Y. Gnedin}

\affil{Institute of Astronomy, Cambridge CB3 0HA, England \\
       ognedin@ast.cam.ac.uk}

\and
\author{Jeremiah P. Ostriker}

\affil{Princeton University Observatory, Princeton, NJ~08544 \\
       jpo@astro.princeton.edu}

\begin{abstract}

The dynamical evolution of galaxies in clusters is modified if dark
matter is self-interacting.  Heat conduction from the hot cluster halo
leads to evaporation of the relatively cooler galactic halos.  The
stellar distribution would adiabatically expand as it readjusts to the
loss of dark matter, reducing the velocity dispersion and increasing the
half-light radius.  If the dark matter content within that radius was
$f_{dm} = 25-50\%$ of the total, as indicated by current observations,
the ellipticals in clusters would be offset from the fundamental plane
relation beyond the observational scatter.  The requirement that their
halos survive for a Hubble time appears to exclude just that range of
the dark matter cross-section, $0.3 \lesssim \sigma_p/m_p \lesssim 10^4$
cm$^2$ g$^{-1}$, thought to be optimal for reducing central halo cusps,
unless $f_{dm} < 15\%$.  If the cross-section is allowed to vary with
the relative velocity of dark matter particles, $\sigma \propto
v^{-2\delta}$, a new problem of evaporation of dark matter arises in the
dwarf galaxies with low velocity dispersion.  The halos of large
galaxies in clusters and dwarf galaxies in the Local Group can both
survive only if $\delta < 1.1$ or $\delta > 1.8$.  In either case the
problem of central density cusps remains.

\end{abstract}

\section{Thermal equilibrium in collisional dark matter halos}

Self-interacting dark matter has been proposed recently to alleviate
some of the problems of galaxy formation arising in the standard cold
dark matter models \cite{SS:00}, \cite{Wetal:00}.  By allowing for heat
conduction into the cooler galactic center, the collisional dark matter
model reduces the density cusp, which is predicted by numerical
simulations in the CDM model \cite{Moore:99}, and seems to be in
conflict with observations.  A drawback of such a scenario is that the
desired heat conduction takes place in all dark matter-dominated
systems, producing undesired phenomena in some situations.  In clusters
of galaxies, the hot cluster halo would tend to evaporate the relatively
cooler galactic halos, giving strict limits on the allowed range of the
self-interaction cross-section.

Depending on the value of the cross-section per particle mass,
$\sigma_p/m_p$, collisional dark matter can act either as a fluid or as
a rarefied gas.  The mean free path of dark matter particles is $\lambda
\approx m_p/(\rho \sigma_p)$, where $\rho$ is the dark matter density.
In the fluid regime, $\lambda(r) \ll r$, the heat diffuses out of the
hot region (cluster halo) to the cool region (galaxy halo) on the
conduction time scale.  In the opposite optically-thin regime, $r <
\lambda$, the interaction is an elastic scattering of the particles.

Numerical simulations in both fluid \cite{Metal:00}; \cite{Yetal:00} and
scattering regimes \cite{B:00}; \cite{KW:00} have shown that in isolated
halos the central cusp is indeed replaced by a soft core, as desired,
for a suitable range of cross-section.  However, a large range of
cross-section appears to be already ruled out by various observations.
The experimental limits and the required rates of core evolution are
estimated to narrow the currently allowed values to the range $0.1 <
\sigma_p/m_p < 6$ cm$^2$ g$^{-1}$ \cite{Wetal:00}.

Theoretical arguments seem to be most powerful for the clusters of
galaxies.  Self-interaction tends to make clusters more spherical than
follows from the lensing observations.  For example, \cite{ME:00}
estimated that this constrains the cross-section to be $\sigma_p/m_p <
0.02$ cm$^2$ g$^{-1}$.  However, there are serious uncertainties
associated with this limit due to statistical and other issues.  In this
paper we consider the effect of thermal evaporation of dark matter halos
in clusters.  Significant evaporation of dark matter halos of the
elliptical galaxies would introduce a systematic offset in their
fundamental plane relations, compared to the field ellipticals.

In \S\ref{sec:fluid} we calculate the dynamics of spherically-symmetric
galaxy halos in the fluid regime solving the diffusion equation.  In
\S\ref{sec:scat} we estimate the evaporation time analytically in the
scattering regime.  In \S\ref{sec:tempdep} we consider the case of a
temperature-dependent cross-section that changes the dynamics of the
inner parts of the halo.  Finally, in \S\ref{sec:ell} we show that the
evaporation of dark matter halos is probably inconsistent with
observational scatter of the fundamental plane relation.

\section{Heat conduction in the fluid regime}
  \label{sec:fluid}

In the fluid regime ($\lambda \ll r$), the frequent interactions of dark
matter particles can be described by a diffusion equation
\begin{equation}
T {dS \over dt}
   = {1 \over n r^2} 
     {\partial \over \partial r} r^2 \kappa {\partial T \over \partial r},
   \label{eq:diff}
\end{equation}
where $S$ is the entropy, $n(r) = \rho(r)/m_p$ is the number density of
dark matter particles of mass $m_p$, and $T$ is the temperature defined
as
\begin{equation}
    {1\over 2} k_B T = {1\over 2} m_p u^2,
\end{equation}
where $u$ is the one-dimensional thermal velocity of dark matter
particles.  The coefficient of thermal conductivity for the monatomic
gas is $\kappa \sim c_v \lambda n u$, where $c_v = 3k_B/2$ is the
specific heat per particle.  From theoretical arguments \cite{HD:00},
the cross-section is either constant or inversely proportional to
temperature squared: $\sigma = \sigma_p \, (T/T_p)^{-\delta}$, $\delta =
0$ or 2.  Also, Wyithe, Turner \& Spergel (2000) have recently proposed
an intermediate case, $\delta = 1/2$.  Therefore, we take
\begin{equation}
  \kappa = {3\over 2} {k_B^{3/2} \, T^{1/2} \over m_p^{1/2} \, \sigma_p}
           \left({T \over T_p}\right)^\delta.
\end{equation}
The entropy $S$ relates to the equation of state $P = n k_B T \propto
\rho^\gamma$ with $\gamma = 5/3$.  It can be written as
\begin{equation}
  T {dS \over dt} = {d \over dt} {3\over 2} k_B T + P {d \over dt} {1\over n}
    = \rho^{\gamma-1} {d \over dt} {3\over 2} {k_B T \over \rho^{\gamma-1}}.
\end{equation}
Assuming that heat conduction is slow compared to the dynamical time in
the galaxy, equation (\ref{eq:diff}) needs to be solved simultaneously
with the hydrostatic equilibrium equation.  Appendix \ref{sec:scheme}
contains the detailed implementation of the code.

As an example, we choose an unspectacular elliptical galaxy NGC 4869 in
the Coma cluster, situated at a distance of $0.119^\circ$ from the
cluster center (220 kpc for the adopted Hubble constant $H_0=65$ km
s$^{-1}$ Mpc$^{-1}$).  Its effective radius is $R_e = 4.2$ kpc and the
half-light luminosity in V band is $L_e = 1.8\times 10^{10}\ L_{\sun}$
\cite{Metal:99}.  The mass-to-light ratio within $R_e$, according to the
fundamental plane relation for such a galaxy, is 8 in solar units
\cite{Getal:01}.  This is about twice the ratio expected for an old
stellar population \cite{BM:98} (later we discuss that the most likely
fraction of dark matter mass within $R_e$ is approximately 50\%), and it
is consistent with the observed stellar velocity dispersion 200 km
s$^{-1}$.  We conservatively take the total mass of the dark matter halo
to be twice the mass contained within the half-light radius; $M_g
= 3\times 10^{11}\ M_{\sun}$.

To set the initial profile for the halo we choose the Hernquist model
extending to the tidal radius $R_{g,t}$:
\begin{equation}
  \rho_g = {M_g R_{g,0} (1+c)^2 \over 2\pi c^2} {1 \over r (r+R_{g,0})^3},
\end{equation}
where $c \equiv R_{g,t}/R_{g,0}$ is the concentration parameter.  We
consider three values of the core radius of the galactic halo, $R_{g,0}
= 2.8, 4, 8$ kpc.  The concentration of the halo may affect its
(thermo)dynamical evolution.  The core radius of about 5 kpc is expected
for a galaxy of such luminosity from the relation of \cite{Getal:01}.

The density profile of the Coma cluster consistent with X-ray data
\cite{H:89} is
\begin{equation}
  \rho_{cl} = {\rho_0 \over (1 + r^2/R_{cl,0}^2)^{3/2}},
\end{equation}
with $\rho_0 = 10^{-25}\ h_{50}^2$ g cm$^{-3} = 2.5\times 10^{-3}\
M_{\sun}$ pc$^{-3}$ and $R_{cl,0} = 290$ kpc.  The average cluster
density at the position of NGC 4869 is $\rho_{cl,\rm av} \equiv 3
M_{cl}(r)/4\pi r^3 = 1.7\times 10^{-3}\ M_{\sun}$ pc$^{-3}$.  This tidal
density determines the truncation radius of the galactic halo,
$R_{g,t}$, via $\rho_{g,\rm av} \approx 2 \, \rho_{cl,\rm av}$ and gives
$R_{g,t} = 28\ (M_g/3\times 10^{11} M_{\sun})^{1/3}$ kpc.

The initial profile for the integration is a sum of the halo and cluster
distributions.  To account for the off-center position of the halo in
the cluster, we have reduced the central density of the cluster profile
to the value of the tidal density, $\rho_0 \rightarrow \rho_{cl,\rm
av}$.  This is justified as the galaxy size is much smaller than the
core radius of the cluster and this provides a correct tidal truncation.
Also, as the galaxy orbits around the cluster center, its tidal radius
and other external boundary conditions may vary.  Using the observed
position in the cluster as the average value is reasonable since the
galaxies are expected to be observed at the location where they spend
most of their time.

%
%

Dimensionless equations can be obtained using the characteristic mass $M
\equiv M_g$ and size $R \equiv R_{g,t}$ of the galaxy halo.  The unit of
temperature is then $T_0 \equiv G M m_p / k_B R$, the unit of velocity
is $v_0 \equiv (GM/R)^{1/2}$, and the unit of time
\begin{equation}
  t_0 \equiv {1 \over 12\pi} \, \tau_t \, t_{\rm dyn},
  \label{eq:t0}
\end{equation}
where $t_{\rm dyn} \equiv (GM/R^3)^{-1/2} \approx 1.3\times 10^8$ yr is
the dynamical time and
\begin{equation}
  \tau_t = \left({\sigma_p \over m_p}\right)
           \left({3M \over 4\pi R^2}\right)
	    \left({T_p \over T_0}\right)^\delta
  \label{eq:tau}
\end{equation}
is the optical depth at the tidal radius.

\begin{figure}
\plotone{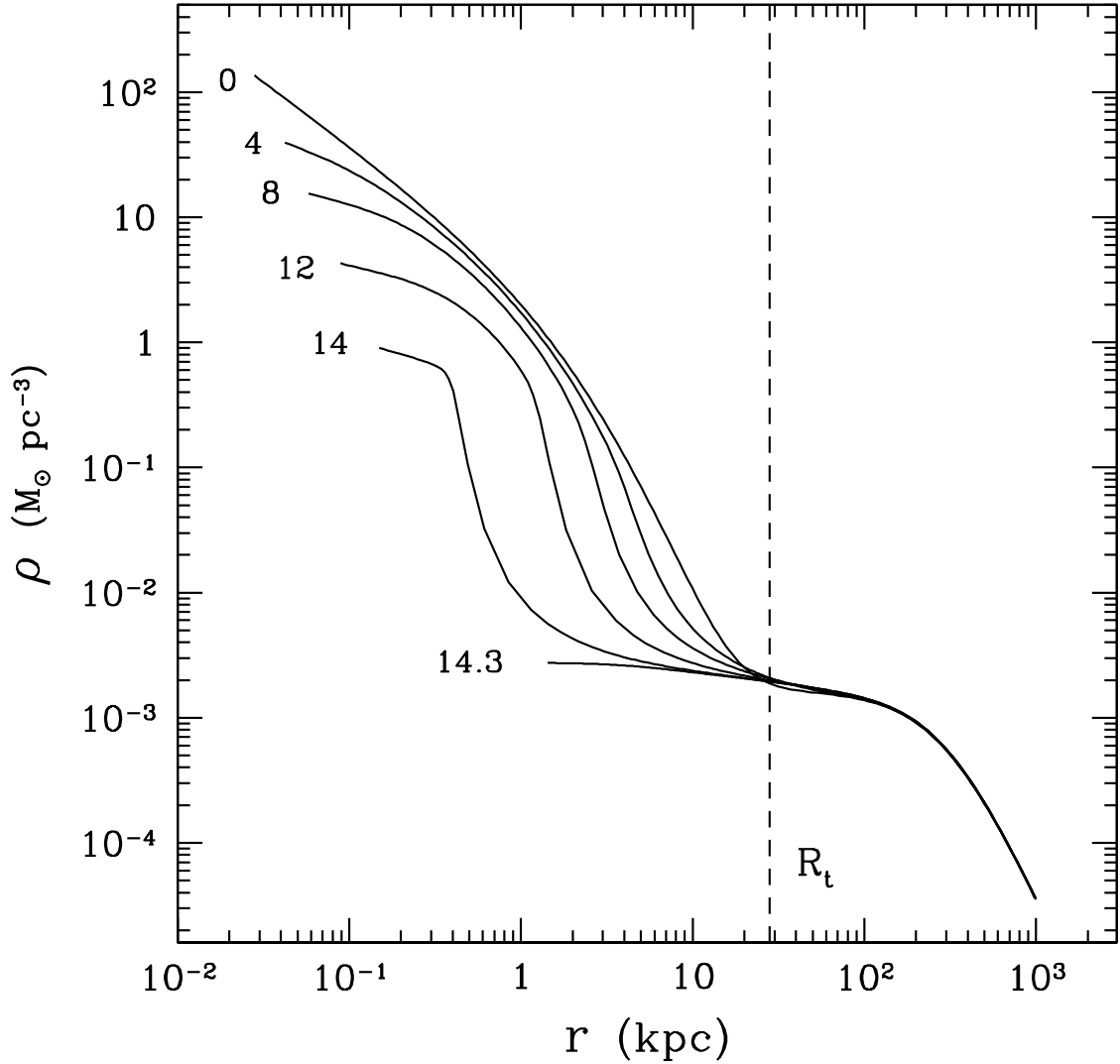}
\caption{Evolution of the density profile of NGC 4869 in the fluid
regime.  The galaxy halo is initially a Hernquist model with the
concentration $c = 7$, the cluster is similar to an isothermal sphere.
The origin of the coordinate system is placed at the galactic center
which does not coincide with the cluster center.  Dashed vertical line
marks the initial tidal radius of the galaxy halo.  The outputs are at
the times indicated, in units of $t_0 = 6.5\times 10^4\, (\sigma_p/m_p)$
yr.  As time progresses the galactic density falls continuously and
approaches the ambient cluster profile.  \label{fig:cl_den}}
\end{figure}

Figure \ref{fig:cl_den} shows the evolution of the galaxy and cluster
halos for the case of constant cross-section, $\delta=0$.  The initial
stage, $t < 4\, t_0$, is driven by internal relaxation in the center but
the subsequent evolution is accelerated by the heat transfer from the
outside.  The rate of evolution in the cluster is faster than in
isolation.  As a result of heating, the galactic halo expands and the
tidal radius shrinks.  Eventually, at $t = 14.3\, t_0$ the whole halo
evaporates leaving the smooth cluster density profile.

\begin{figure}
\plotone{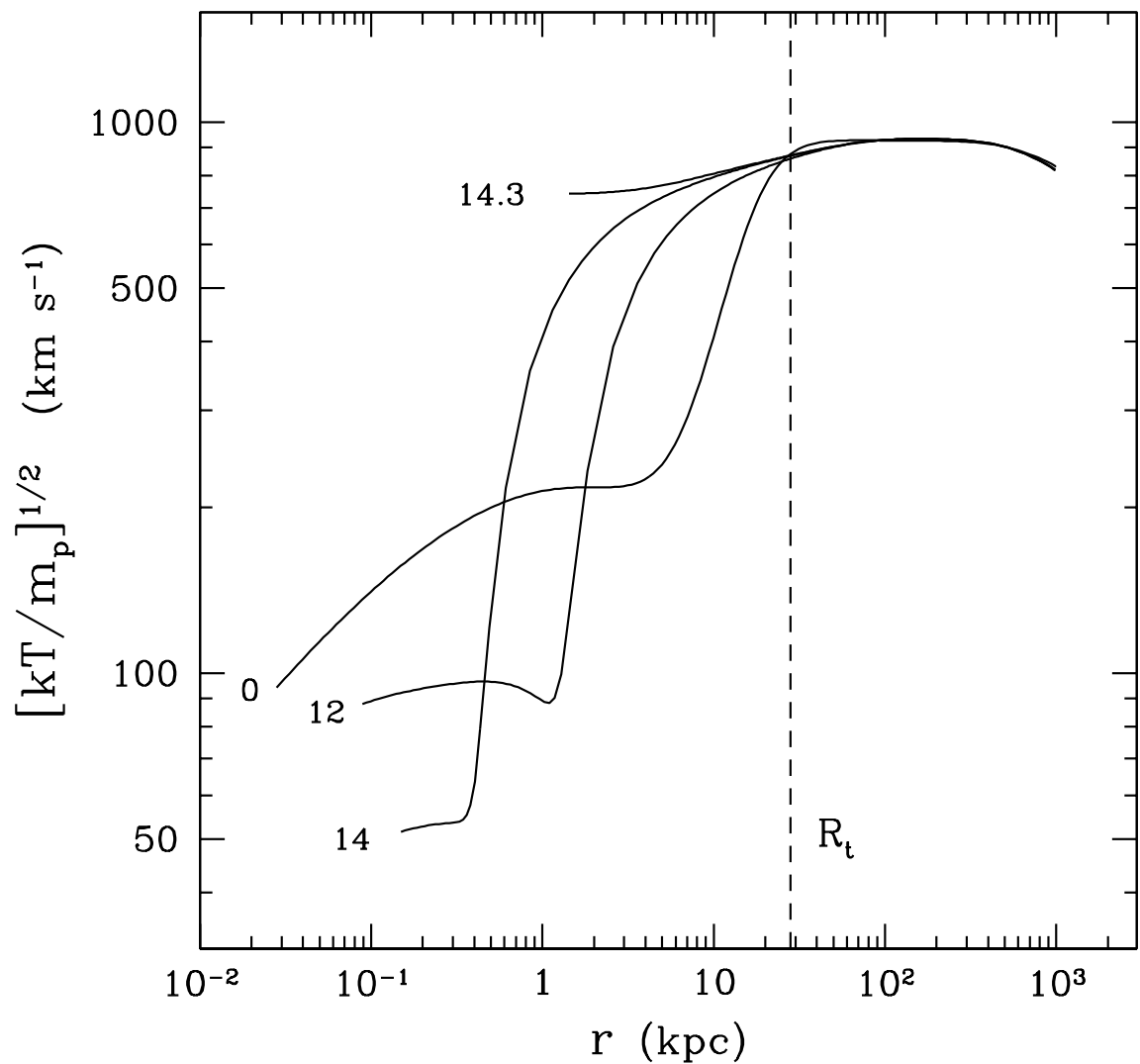}
\caption{Evolution of the temperature profile of NGC 4869 in the fluid
regime.  The temperature is converted to the velocity units,
independent of the particle mass $m_p$.
\label{fig:cl_temp}}
\end{figure}

However, the evolution of the temperature profile is not monotonic in
time (Figure \ref{fig:cl_temp}).  First, the central temperature rises
as the heat is conducted inward from the neighboring regions.  At the
same time, the tidal boundary between the relatively hot and cold
regions moves inward, the halo expands, the confining pressure drops,
and as a result the temperature decreases between 1 and 10 kpc.  Then,
at $t = 12\, t_0$ the central temperature begins to fall as the whole
galaxy becomes isothermal.  At $t = 14\, t_0$ the central temperature
reaches its minimum, even below the initial value.  At that point the
central density is already 100 times lower than initially.  Then, the
remains of the galaxy halo quickly evaporate, and at $t = 14.3\, t_0$
its density falls by another two orders of magnitude, while the
temperature jumps by two orders of magnitude to match the cluster
profile.

Varying the concentration of the galaxy halo does not change its
evolution qualitatively.  A lower density model ($c=3.5$, $R_{g,0} = 8$
kpc, same $R_{g,t}$) disrupts in $9.9\, t_0$, while the higher density
model ($c=10$, $R_{g,0} = 2.8$ kpc) takes about $17\, t_0$.  Thus, the
evaporation time is weakly dependent on the internal structure of the
halo, and the best value can be taken as $14\, t_0$.

The evolution is identical in dimensionless units as long as the
galactic halo is in the fluid regime.  The dark matter cross-section
enters only the unit of time, $t_0 \propto \sigma_p/m_p$.  The larger
the cross-section the shorter is the mean free path and the longer is
the evaporation timescale,
\begin{equation}
  t_{fl} = 14\, t_0
         = 9\times 10^5 \; \left({\sigma_p \over m_p}\right)
           \left({M \over 3\times 10^{11}\, M_{\sun}}\right)^{1/2}
           \left({R \over 28\ \mbox{kpc}}\right)^{-1/2}
           \ \mbox{yr}.
    \label{eq:tfl}
\end{equation}

The stars in the galaxy increase the lifetime of the dark halo by
providing an additional gravitational attraction in the inner parts.
Equation (\ref{eq:tfl}) shows that $t_{fl}$ depends essentially on the
average velocity dispersion of the halo.  Adding the fraction of stellar
mass, $\lesssim 50\%$, would therefore increase the evaporation time by
about 20\%.

Note, that in the fluid regime the density of the cluster particles does
not affect the halo evaporation timescale, as long as the cluster
represents a much larger heat reservoir.  The inward propagation of the
heat wave is governed primarily by the temperature gradient between the
galaxy and the cluster.  If that gradient is always large, the
evaporation time is determined by the internal conduction time and is
insensitive to the cluster parameters.

\section{Evaporation in the scattering regime}
  \label{sec:scat}

The situation is different when the particle mean free path is larger
than $r$: in this case the fluid equation (\ref{eq:diff}) does not
apply.  The critical density of the fluid regime, i.e. the average
density for which the optical depth is unity, is
\begin{equation}
  \rho_{fl} = {m_p \over \sigma_p \, r}.
\end{equation}
For $\sigma_p/m_p \lesssim 50$ cm$^2$ g$^{-1}$, the galactic and cluster
halo densities fall below the critical density.  If the cross-section is
considerably smaller, the evolution in the optically-thin regime can be
calculated at three levels of approximation.

(i) Cluster particles, which have much higher kinetic energy, would
impart a fraction of this energy to the relatively cooler galactic
particles with each collision.  For the same particle mass, the energy
exchange would suffice to unbind the galactic particle.  Thus, each
collision with the cluster particle removes one from the halo.  The mean
free path of the cluster particles through the galactic halo is $\lambda
= m_p/(\sigma_p \, \rho_g)$, and the probability of collision in one
crossing time is $r/\lambda(r)$.  In the optically-thin limit, the
collision time per particle at the tidal radius $R$ is then
\begin{equation}
  t_{\rm coll} \sim t_{\rm cross}(R) \; {\lambda(R) \over R},
\end{equation}
where the crossing time of the cluster particles is $t_{\rm cross} \sim
R/v_{cl} \sim (v_g/v_{cl}) \, t_{\rm dyn}$.  The collision time for all
galactic particles is $t_{sc} = (\rho_g/\rho_{cl}) \, t_{\rm coll}$.
But since the densities of the galaxy and the cluster are similar at the
tidal radius, the timescale is of the same order.

(ii) The calculation can be done more accurately, noting that at any
position within the halo the rate of evaporation is $\dot{\rho} =
-\rho(r)/t_{sc}$.  The amount of bound mass decreases faster because the
tidal boundary shrinks as mass is being lost.  The resulting mass loss
is exponential with the timescale roughly $t_{sc}/2$, depending on the
steepness of the density profile.  With sufficient accuracy we can still
take the evaporation time to be $t_{sc}$.

(iii) Finally, a heat transfer equation can be used, similar to equation
(\ref{eq:diff}), treating the galaxy and cluster particles as two
separate fluids.  In this case the diffusion coefficient has to be
modified to account for the large mean free path of the particles.  The
details are described in Appendix \ref{sec:opthin}.

In the scattering regime, the smaller the cross-section the longer is
the evolution timescale.  At the position of NGC 4869 the cluster
density is $\rho_{cl} \approx 1.3\times 10^{-3}\, M_{\sun}$ pc$^{-3}$
and $v_{cl} \approx 10^3$ km s$^{-1}$.  This gives for the evaporation
time
\begin{equation}
  t_{sc} = 3.5\times 10^9 \; \left({\sigma_p \over m_p}\right)^{-1}
           \left({v_{cl} \over 10^3\ \mbox{km s}^{-1}}\right)^{-1}
           \left({\rho_{cl} \over 1.3\times 10^{-3}\,
                 M_{\sun}\, \mbox{pc}^{-3}}\right)^{-1}
           \ \mbox{yr}.
    \label{eq:tsc}
\end{equation}

\begin{figure}
\plotone{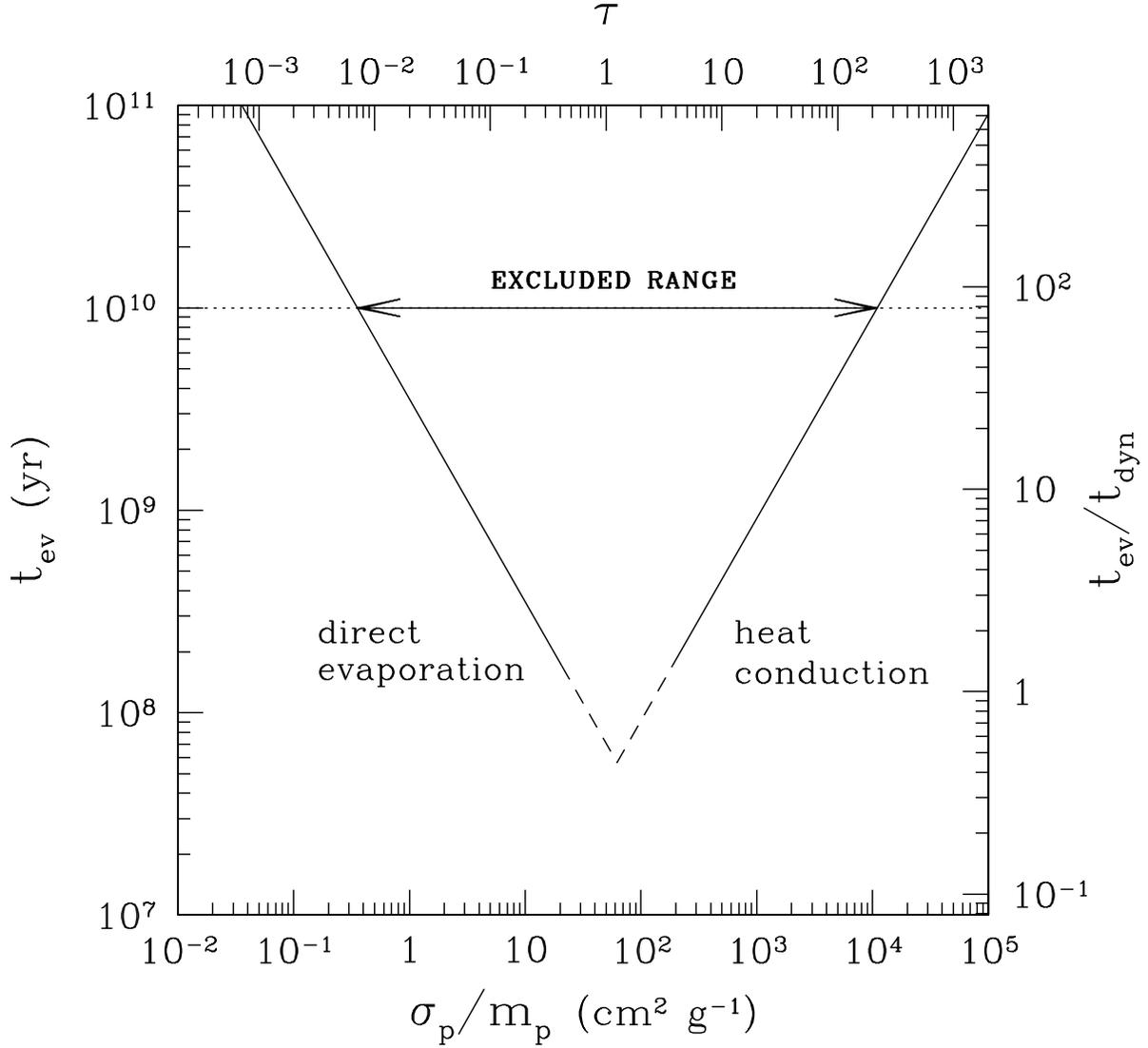}
\caption{Evaporation time of the dark matter halos of elliptical
galaxies in clusters in the fluid ($\tau_t > 1$) and scattering ($\tau_t
< 1$) regimes.  The lines broke into dashes in the marginal range, for
the optical depth at the tidal radius $1/3 < \tau_t < 3$.  The arrows
indicate the range of the cross-section excluded by the requirement that
the halos survive for $10^{10}$ yr.  
  \label{fig:cross}}
\end{figure}

Figure \ref{fig:cross} shows the evaporation time in the fluid and
optically-thin regimes.  The fastest evolution is achieved for the
marginal case $\lambda \approx R$, in which domain we have not made a
calculation.  This intermediate region is sketched in with the dashed
line.  There, the estimates of the evaporation time should agree in the
two regimes.  For the chosen tidal density, the critical value of the
cross-section is $\sigma_p/m_p = 50$ cm$^2$ g$^{-1}$.  The two time
scales, $t_{fl} = 4.5\times 10^7$ yr and $t_{sc} = 7.0\times 10^7$ yr,
agree to within a factor of two.

\section{Other dynamical effects}
  \label{sec:other}

Another effect can be important in the fluid regime (pointed out by the
referee, Andi Burkert).  Ram pressure of the cluster halo strips outer
parts of the galactic halo as the latter moves around the cluster.  The
restoring gravitational force of the stars balances the ram pressure
force on dark matter particles only at two half-light radii of NGC 4869.
At the tidal radius the gravitational force falls to 5\% of that.  Thus,
we expect that on a dynamical timescale the material outside about $2\,
R_e \approx 8$ kpc from the galactic center would be stripped.  This
would only accelerate heat conduction to the inner region and accelerate
evaporation of the halo.

In the optically-thin regime, ram pressure is equivalent to the direct
scattering of particles and does not add to a new effect.  The orbital
motion of the halo contributes to the relative velocity between the
cluster and galactic particles, which is of the same order.  Thus, the
constraint on the cross-section can be strengthened by a factor
$\sqrt{2}$ (cf eq. [\ref{eq:tsc}]).

Dynamical friction of the galactic orbit in the cluster would also speed
up evaporation in either regime.  As the halo gradually spirals in
towards the cluster center, it encounters a higher density of cluster
particles with higher temperature.  However, we ignore this effect as it
is unlikely to change the constraint on the interaction cross-section
significantly.

\section{Constraints from the fundamental plane relation}
  \label{sec:ell}

Elliptical galaxies satisfy a tight fundamental plane relation
\cite{DD:87}.  One of its projections, the Faber-Jackson relation, has
an observed scatter in the central velocity dispersion $\Delta
\log{\sigma_c} < 0.07$ both in optical bands and in the near-infrared
\cite{Metal:99}.  If the ellipticals lose their dark matter halos, their
stellar distributions would adiabatically expand and reduce the central
dispersion, while conserving the total luminosity.  This would introduce
a systematic offset in the Faber-Jackson relation, compared to the field
ellipticals that have not lost their dark matter halos.

The adiabatic expansion is characterized by conservation of the
azimuthal action, $G M(r) r$.  A spherical shell initially at radius $r$
would adiabatically expand to the radius $r'$ such that
\begin{equation}
  M_*(r') \, r' = [M_{dm}(r) + M_*(r)] \, r.
\end{equation}
The amount of expansion depends on the fraction of dark matter within
$r$: $f_{dm}(r) \equiv M_{dm}(r)/M_{tot}(r)$.  
The surface brightness of elliptical galaxies is well fitted by the de
Vaucouleurs profile.  The effective radius, $R_e$, characterizes the core
of the stellar distribution and contains half the luminous mass,
$M_*(R_e) = M_*/2$.  As a result of adiabatic expansion of the core, the
effective radius would rise by a factor
\begin{equation}
  {R'_e \over R_e} = {1 \over 1 - f_{dm}(R_e)}.
  \label{eq:rere}
\end{equation}

As the core expands, the stellar velocity dispersion falls.  If the
dispersion profile is flat in the core, i.e. $\sigma_c$ is determined by
the mass distribution within $R_e$, the dispersion would drop by $\Delta
\log{\sigma_c} = -\log{R'_e/R_e}$.  Then, the galaxy would still satisfy
the Faber-Jackson relation within the errors if the dark matter fraction
is $f_{dm}(R_e) < 1 - 10^{-\Delta \log{\sigma_c}} = 0.15$.

Spectroscopic observations of elliptical galaxies usually have apertures
of approximately 3\arcsec, corresponding to 1.5 kpc at the distance of
the Coma cluster.  The aperture size may be small enough compared to
$R_e$ so that it contains only a small fraction of dark matter, and the
measured dispersion would respond more weakly to the expansion of the
core.  As an alternate limiting case, we consider that $\sigma_c$ may
not change at all, even though the surface brightness does decrease
within $R_e$.

In addition to the change of the Faber-Jackson relation, expansion of
the stellar core would lead to a shift away from the fundamental plane
relation.  The latter can be expressed as the variation with mass of the
mass-to-light ratio within $R_e$: $M_e/L_e \propto M_e^{0.184}$
\cite{Betal:97}, where the mass is defined as $M_e \propto \sigma_c^2 \,
R_e$.  The expansion of the core would change the measured value of
$M_e$ while the integrated luminosity $L_e$ remains the same.

\begin{figure}
\plotone{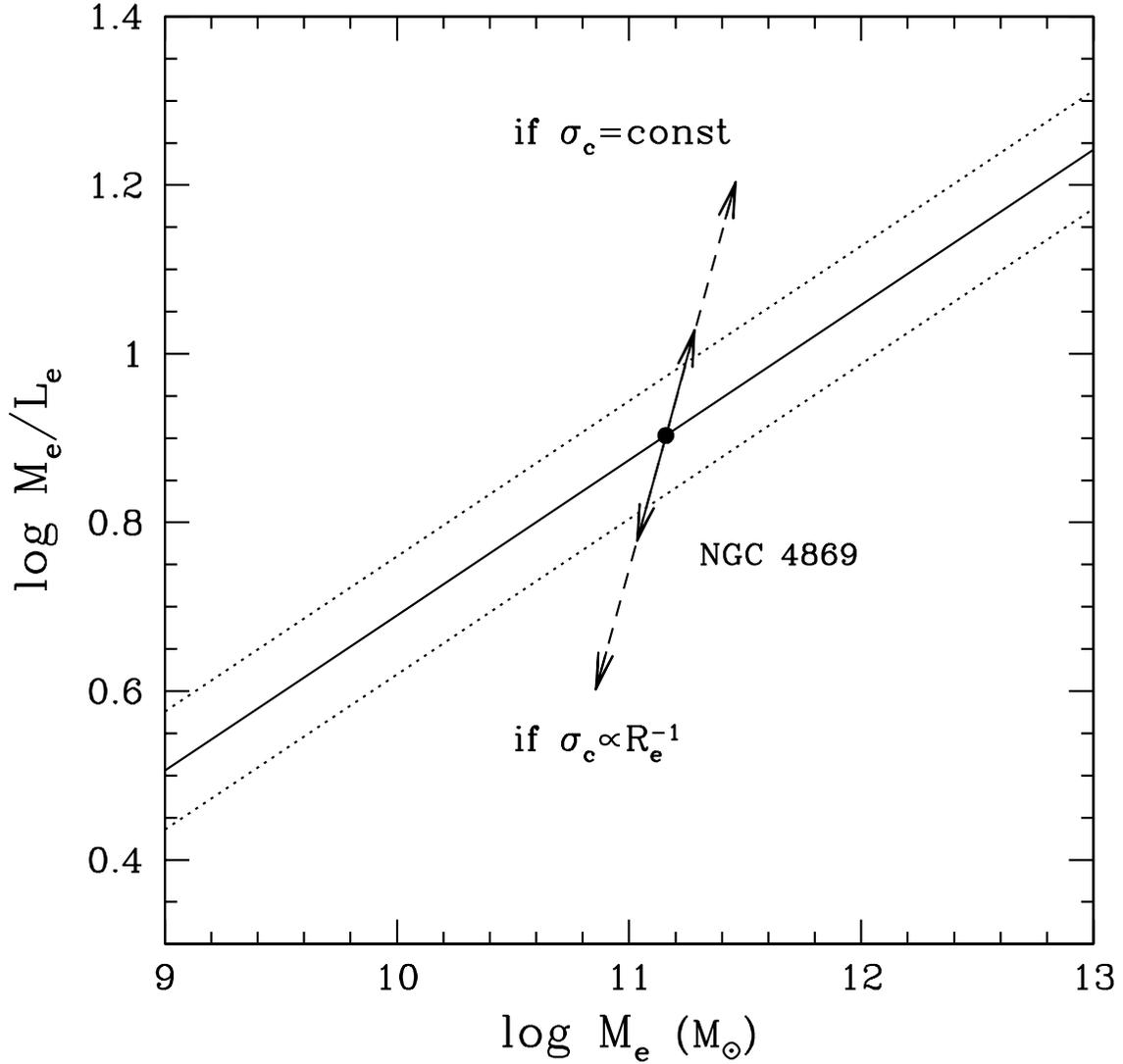}
\caption{Fundamental plane relation for the elliptical galaxies (solid
line) and its scatter (dots).  The half-light mass-to-light ratio
$M_e/L_e$ is in solar units.  If NGC 4869 loses its dark matter halo, it
would deviate from the plane either upward (if the core velocity
dispersion stays constant as the effective radius expands) or downward
(if $\sigma_c \propto R_e^{-1}$).  The length of solid arrows is
calculated using $f_{dm}=0.25$, a minimum dark matter fraction within
$R_e$, while dashed arrows are for best-fitting model, $f_{dm}=0.5$.
  \label{fig:fp}}
\end{figure}

Figure \ref{fig:fp} demonstrates the effect for NGC 4869.  If the core
velocity dispersion falls as the effective radius expands, $\sigma_c
\propto R_e^{-1}$, both the mass and the mass-to-light ratio decrease by
the factor $(1 - f_{dm})^{-1}$ (eq. [\ref{eq:rere}]).  If the velocity
dispersion does not respond to the core variation at all, both variables
increase by the same amount, and the galaxy again moves away from the
fundamental plane.  The magnitude of the offset is determined by the
dark matter fraction $f_{dm}(R_e)$.

The amount of dark matter in elliptical galaxies is harder to measure
than in spiral galaxies.  Fitting results of the X-ray profiles of 42
bright elliptical galaxies \cite{LW:99} require $f_{dm}(R_e) \ge 0.2$ to
explain the spectral parameter $\beta = 0.5$.  The best fitting model
with $M_{dm} = 3 M_*$ has $f_{dm}(R_e) \approx 0.5$.  This estimate,
that dark matter contributes 50\% of the total mass of an elliptical
galaxy at the half-light radius, is consistent with our model of NGC
4869, and also with the model of NGC 4472 obtained using the velocity
dispersion of globular clusters \cite{Zetal:00}.  We use the value
$f_{dm} = 0.5 \pm 0.25$ in Figure \ref{fig:fp}.

%

If the dark matter fraction $f_{dm}(R_e) > 0.15$, elliptical galaxies in
clusters need to retain their dark matter halos in order to stay on the
same fundamental plane as the field ellipticals.  The fundamental plane
relation is observed to be the same in clusters and in the field, and
within clusters there is no differential radial effect.  The evaporation
time of the halos, in either fluid or scattering regime, must be longer
than the Hubble time.  Figure \ref{fig:cross} shows that this excludes
the range of the collisional cross-section $0.35 < \sigma_p/m_p <
1.1\times 10^4$ cm$^2$ g$^{-1}$.

Note, that similar constraints apply to other types of galaxy.  The
evaporation of dark matter halos of spiral galaxies in clusters would
introduce a systematic offset in the Tully-Fisher relation.  In
addition, bulges of spiral galaxies are observed to lie on the same
fundamental plane as elliptical galaxies.  Expansion of the cores of
elliptical galaxies in clusters would break this law.

\section{Temperature-dependent cross-section}
  \label{sec:tempdep}

The evaporation and heat conduction are modified if the self-interaction
cross-section varies with the relative velocity of dark matter
particles.  If self-interaction is coupling-limited \\
\cite{HD:00}, the
cross-section decreases with temperature as $\sigma = \sigma_p \,
(T/T_p)^{-2}$.  The expected temperature scale for likely candidate
particles $T_p \ga 1$ keV.  Other forms of interaction are conceivable
and therefore we consider a general function for the cross-section
varying as $\sigma = \sigma_p \, (T/T_p)^{-\delta}$.  Obviously, only
the product $\sigma_p \, T_p^\delta$ can be constrained.

Note that different temperatures determine the cross-section depending
on the optical depth.  If $\tau > 1$, it is the local temperature of
the colder halo; if $\tau < 1$, it is the temperature of the hotter
halo, since it contributes mainly to the relative particle velocity.
The optical depth itself depends on the temperature, but at the tidal
radius the two regimes coincide as $T_{\rm cold} \rightarrow T_{\rm
hot}$ for $\tau > 1$.

The most optimistic values of the cross-section, as not being in
contradiction with observations described in the previous section, lie
in the optically-thin regime with $\sigma_p/m_p \sim 0.1$ cm$^2$
g$^{-1}$.  Since we have the freedom in choosing $T_p$, we fix the
temperature scale at $u_p \equiv (k_B T_p/m_p)^{1/2} = 10^3$ km
s$^{-1}$.  In this case the evaporation time in the optically-thin
regime remains the same:
\begin{equation}
  t_{sc} = 3.5\times 10^9 \; \left({\sigma_p \over m_p}\right)^{-1}
           \left({v_{\rm hot} \over 10^3\ \mbox{km s}^{-1}}\right)^{-1+2\delta}
           \left({\rho_{\rm hot} \over 1.3\times 10^{-3}\,
                 M_{\sun}\, \mbox{pc}^{-3}}\right)^{-1}
           \ \mbox{yr},
    \label{eq:tsc2}
\end{equation}
where $v_{\rm hot} = v_{cl}$ and $\rho_{\rm hot} = \rho_{cl}$
correspond to the hotter halo.  The optical depth at the tidal radius
is also constant regardless of $\delta$; $\tau_t = 0.02\,
\sigma_p/m_p$.  Throughout this section, the cross-section $\sigma_p$
refers to the particular choice of $u_p$.

In the fluid regime, the diffusion equation is different with $\delta >
0$ (see eq. [\ref{eq:fluid}]).  Compared with the constant cross-section
case, heat conduction is suppressed in the cooler central parts of the
galaxy while the heat wave from the cluster propagates faster in the
outer parts.  Thus, for $\delta > 0$, the desired central heating is
reduced, but the unwanted evaporation of the outer region is increased.

This requires modification of the external boundary condition in the
numerical code.  The simplest solution is to neglect the scattering
events in the cluster halo and move the external boundary to the
galactic tidal radius.  Thus, the galaxy is bound by a fixed cluster
pressure at $R$.  The evolution proceeds similarly to the case
described in \S\ref{sec:fluid}, but somewhat faster in the outer
parts.  The evaporation time is $t_{fl} = q(\delta) \, t_0(\delta)$,
where $q(\delta)$ is decreasing with increasing $\delta$; $q(\delta)
\approx 1 - 10$.  However, the unit of time $t_0$ is scaled up with
$\delta$.  Thus, the evaporation time in the fluid regime is
\begin{equation}
  t_{fl} = 3.0\times 10^5 \; q(\delta) \; \left({\sigma_p \over m_p}\right)
           \left({v_{\rm cold} \over 10^3\ \mbox{km s}^{-1}}\right)^{1-2\delta}
           \ \mbox{yr},
    \label{eq:tfl2}
\end{equation}
where $v_{\rm cold} \equiv (GM/R)^{1/2}$ is the unit of velocity of
the colder halo.  For $\delta=2$ and $v_{\rm cold} = 215$ km s$^{-1}$,
appropriate for NGC 4869, $t_{fl} \sim 3\times 10^7 \, (\sigma_p/m_p)$
yr.  The fluid regime applies to NGC 4869 for $\sigma_p/m_p > 50$
cm$^2$ g$^{-1}$.

Even if we take $\sigma_p/m_p = 0.1$ cm$^2$ g$^{-1}$, so that the
optical depth is very low at the tidal radius of NGC 4869, the central
part may become optically-thick if $\delta > 0$.  Because of the lower
velocity dispersion, the cross-section would rise steeply there.  For
example, for $\delta = 2$, the inner 8 kpc are optically-thick.  Because
of the negative heat capacity of the self-gravitating central region,
heat transfer into the center would only lower the temperature and
increase the cross-section.  A natural lower limit of the temperature is
the velocity dispersion provided by the stars, of order 200 km s$^{-1}$.
The central density of dark matter would fall until the optical depth
falls below unity.  If we require that the dark matter mass within $r_s$
does not fall by more than 50\%, the cross-section must not decrease
with temperature faster than $\delta \approx 1.1$.

Note that for $\delta > 0$ the optical depth increases for smaller
halos, by a factor $(u_p/v_{\rm hot})^{2\delta}$ (cf
eq. [\ref{eq:tau}]).  In the optically-thin regime, the cross-section
increases by the same factor.  Therefore, the halos of dwarf galaxies
would evolve faster within the Local Group ($v_{\rm hot} \sim
220/\sqrt{2} \approx 150$ km s$^{-1}$) than within large clusters
($v_{\rm hot} \sim 1000$ km s$^{-1}$).  For example, Carina, Draco,
and Ursa Minor are entirely dark-matter dominated dwarf galaxies, with
the central velocity dispersion $v_{\rm cold} \approx 6.6$ km s$^{-1}$
\cite{M:98}.  If the dark matter particles in their halos become
optically-thick to self-interaction the cross-section grows even
further, by a factor $(u_p/v_{\rm cold})^{2\delta}$ compared to the
$\delta=0$ case.  For the large or small enough values of $\delta$,
depending on the regime, the dwarf halos would completely evaporate.
This allows us to constrain a possible range of $\delta$ from the
requirement that the dwarf halos survive for the Hubble time.

The density of the Galaxy halo at a distance of 70-100 kpc, where the
dwarf spheroidals are located, is of order $\rho_{\rm hot} \sim
10^{-4}\, M_{\sun}\, \mbox{pc}^{-3}$.  The estimated masses of the dwarf
halos are $M \approx 2\times 10^7\, M_{\sun}$, and the corresponding
tidal radii $R \approx 2$ kpc.  We shall fix the normalization of the
cross-section at $\sigma_p/m_p = 0.1$ cm$^2$ g$^{-1}$, suitable for
large galaxies.  Then, the dwarf halos would satisfy the condition
$\tau_t < 1$ as long as $\delta < 2.8$.  In fact, in the cores of these
galaxies the density is much higher than average, $\rho_0 \sim 0.4\,
M_{\sun}\, \mbox{pc}^{-3}$ within the core radii $r_0 \sim 200$ pc
\cite{M:98}.  In order for the centers to remain in the optically-thin
regime, the halos need $\delta < 1.7$.  From equation (\ref{eq:tsc2}),
the evaporation time is longer than $10^{10}$ yr if $\delta < 1.5$.
Thus, in the scattering regime $\delta$ should be small enough to rescue
the dwarfs.\footnotemark

\footnotetext{
These constraints become more severe if we take a larger normalization
of the cross-section, $\sigma_p/m_p = 1$ cm$^2$ g$^{-1}$.  In this
case, the cores of the dwarfs remain optically-thin for $\delta < 1.1$
and survive for the Hubble time if $\delta < 0.9$.
}

If the slope $\delta$ is high enough ($\delta > 1.7$), the dwarf
galaxies may fall into the optically-thick regime even for
$\sigma_p/m_p = 0.1$ cm$^2$ g$^{-1}$.  From equation (\ref{eq:tfl2}),
the evaporation time is longer than $10^{10}$ yr if $\delta > 1.8$ (or
$\delta > 1.5$ for $\sigma_p/m_p = 1$ cm$^2$ g$^{-1}$).  In other
words, the cross-section should increase substantially enough that
heat conduction slows down.

Thus, the presence of the dark matter-dominated dwarf galaxies in the
Local Group rules out a certain range of the temperature slope, $1.5 <
\delta < 1.8$, even for the most optimistic value of $\sigma_p/m_p =
0.1$ cm$^2$ g$^{-1}$.  Together with the constraint from NGC 4869, this
excludes the values $1.1 < \delta < 1.8$.

\section{Conclusions}

We have considered astrophysical constraints on the possible
cross-section of self-interacting dark matter.  The fundamental plane
relation of elliptical galaxies requires that their dark halos largely
survive the heating from the hot cluster halos.  This condition applies
to galaxies with the half-light dark matter fraction $f_{dm} > 0.15$.
For a velocity-independent cross-section, the evaporation time of such
halos is shorter than the Hubble time if $0.3 \lesssim \sigma_p/m_p
\lesssim 10^4$ cm$^2$ g$^{-1}$.

The parameter space can be enlarged by assuming $\sigma_p/m_p = 0.1$
cm$^2$ g$^{-1}$ and allowing for a velocity dependence of the
cross-section.  The large halos would be unaffected, as long as they
remain in the optically-thin regime, but the dwarf halos within larger
halos would evaporate if the optical depth rises towards unity.  The
presence of dwarf spheroidals in the Local Group excludes the range of
the power-law slope $1.1 < \delta < 1.8$.

Combined with other constraints in the literature \cite{Wetal:00}, this
appears to exclude the most interesting range of the cross-section for
the purpose of lowering the core densities of galactic halos.  In order
to reduce the central density cusp by scattering, the optical depth at
the scale radius of the halo $R_{g,0}$ should be close to unity
\cite{SS:00}.  For a Hernquist model of the halo, $\tau(R_{g,0})$
exceeds the optical depth at the tidal radius, $\tau_t$, by a factor
$(1+c)^2/4 \sim 10 - 30$.  We can therefore expect the optimum regime to
be $\tau_t \sim 0.03 - 0.1$, but as Figure \ref{fig:cross} demonstrates,
this regime would lead to evaporation of the halo in less than a Hubble
time.

Thus, the original motivation for considering self-interacting dark
matter may require collisional cross-sections too large to be consistent
with observations of galaxies in clusters.  However, the cross-section
in the optically-thin regime ($\sigma_p/m_p \sim 10^{-1}$ cm$^2$
g$^{-1}$) may still have very interesting consequences for the growth of
central black holes \cite{O:00}.

\acknowledgements

We would like to thank Martin Rees, David Spergel, and Paul Steinhardt
for valuable discussions, Nick Gnedin for advice with the numerical
scheme, and the referee Andi Burkert for useful comments.  This work was
supported in part by PPARC and by NSF grants ASC-9740300 and
AST-9803137.

\appendix

\section{Numerical scheme for the fluid regime}
 \label{sec:scheme}

For computational convenience, the equations are rewritten in Lagrangian
form using the new variable of integration $dM_r = 4\pi \rho r^2 dr$.
In dimensionless form they become
\begin{eqnarray}
&& {\partial r \over \partial M_r} 
   = {1 \over 4\pi \rho r^2}, \\
&& {\partial P \over \partial M_r}
   = - {M_r \over 4\pi r^4}, \\
&& \rho^{\gamma-1} {\partial \over \partial t} {T \over \rho^{\gamma-1}}
   = {\partial \over \partial M_r}
   \left( r^4 \rho T^{\delta+1/2} {\partial T\over \partial M_r} \right).
   \label{eq:fluid}
\end{eqnarray}
The unit of time (eq. [\ref{eq:t0}]) absorbs all parameters of the
problem.  The equation of state is enforced by defining the specific
entropy as $P = s \rho^\gamma$.  The central differencing scheme is
\begin{eqnarray}
&& {r_{i+1}^{3-\alpha_i} - r_i^{3-\alpha_i} \over m_{i+1}-m_i} 
         - {3-\alpha_i \over 4\pi} {1\over 2}
           \left({1\over \rho_i r_i^{\alpha_i}}+
            {1\over \rho_{i+1} r_{i+1}^{\alpha_i}}\right) = 0,
   \label{eq:r_diff} \\
&& s_i \rho_i^\gamma - s_{i+1} \rho_{i+1}^\gamma
         - {m_{i+1}-m_i \over 4\pi} {1\over 2}
           \left({m_i \over r_i^4} + {m_{i+1} \over r_{i+1}^4}\right) = 0,
   \label{eq:rhoT_diff} \\
&& {T_i - T_i^* \over \Delta t} \equiv \Gamma_i =
     {F_{i+1/2} - F_{i-1/2} \over (m_{i+1}-m_{i-1})/2},
   \label{eq:temp_diff}
\end{eqnarray}
where $T^*$ is the solution at the previous time step, and
\begin{equation}
  F_{i+1/2} = {1\over 2} \left( r_i^4 \rho_i T_i^{\delta+1/2} +
        r_{i+1}^4 \rho_{i+1} T_{i+1}^{\delta+1/2} \right)
        {T_{i+1} - T_i \over m_{i+1}-m_i},
\end{equation}
\begin{equation}
  \alpha_i \equiv -{\ln{\rho_{i+1}}-\ln{\rho_i} \over
                    \ln{r_{i+1}}-\ln{r_i}}.
\end{equation}
The time-variable local slopes of the density distribution, $\alpha_i$,
provide the most accurate integration.  The grid is logarithmic in
radius.  The inner boundary condition assures constant mass of the first
mesh point, $\rho_1 r_1^3 = const$.  The outer boundary condition fixes
the external pressure of the continuous cluster distribution, $\rho_N
T_N = const$.

The finite difference equations are solved using the iterative
correction procedure.  Corrections to the previous configuration are
sought: $\delta r \equiv r-r^*$, $\delta \rho \equiv \rho-\rho^*$,
$\delta T \equiv T-T^*$.  First, the temperature is updated using
equation (\ref{eq:temp_diff}).  Then the entropy is fixed,
\begin{equation}
  s_i \equiv {T_i \over (\rho_i^*)^{\gamma-1}},
\end{equation}
and $\delta \rho_i$ are calculated from $i=N$ downward and $\delta r_i$
are calculated from $i=1$ upward.  The corrections are added to the
previous solution and equations (\ref{eq:r_diff},\ref{eq:rhoT_diff}) are
iterated until the corrections converge.  The time step is chosen such
that the relative corrections $\delta r/r$, $\delta \rho/\rho$, $\delta
T/T$ do not exceed 10\%.

\section{Heat transfer in the optically-thin regime}
  \label{sec:opthin}

The coefficient of thermal conductivity for the monatomic gas is $\kappa
\approx c_v n \lambda^2/t_{\rm coll}$, where $c_v = 3k_B/2$ is the
specific heat per particle.  In the optically-thick regime, $\lambda
\approx u t_{\rm coll}$ and $\kappa = c_v \lambda n u$.  In the
optically-thin regime, $\lambda(r) > r$, this should be modified to
replace $\lambda$, which is now larger then the size of the system, with
$u t_{\rm dyn}$, the characteristic distance traveled by the particles
(similar to heat conductivity in globular clusters; \cite{LBE:80}).
Thus, the conductivity is multiplied by the local optical depth $\tau(r)
\equiv t_{\rm dyn}(r)/t_{\rm coll}(r)$ squared.  Taking $t_{\rm dyn} =
(4\pi G \rho)^{-1/2}$ and $t_{\rm coll} = m_p/(\rho \sigma_p u)$, we
have
\begin{equation}
  \kappa_o = {3\over 2} {k_B^{3/2} \, T^{1/2} \over m_p^{1/2} \, \sigma_p}
           \times \left({\sigma_p \over m_p}\right)^2
           {k_B \, T \, \rho \over 4\pi G \, m_p}.
\end{equation}

For simplicity, we assume the density and temperature of the cluster
particles to be fixed and uniform throughout the halo.  The change of
the entropy of the galactic halo particles is the sum of the internal
heat conduction with other halo particles and the direct heating by
cluster particles.
\begin{equation}
  T {dS \over dt} = {dQ_{\rm cond} \over dt} + {dQ_{\rm heat} \over dt}
     = {1 \over n r^2} 
       {\partial \over \partial r} r^2 \kappa_o {\partial T \over \partial r}
     + c_v {T_{cl} - T \over t_{sc}}.
\end{equation}
With a new unit of time, $t_0 \equiv 3/(64\pi^3 \tau_t) \, t_{\rm dyn}$,
the equation can be written in dimensionless form
\begin{equation}
   \rho^{\gamma-1} {\partial \over \partial t} {T \over \rho^{\gamma-1}}
   = {\partial \over \partial M_r}
      \left( {1 \over 4\pi} r^4 \rho^2 T^{\delta+3/2}
             {\partial T\over \partial M_r} \right)
   + {T_{cl} - T \over t'_{sc}},
\end{equation}
where $t'_{sc} \equiv 16\pi^2 / (\rho_{cl} v_{cl})$.



\end{document}